\documentstyle[preprint,aps,floats,epsfig]{revtex}

\begin{document}
%\preprint{CERN-TH/2001-??}
\preprint{Stanford ITP-01/36}
\draft
\tighten

\title{String Balls at the LHC\\ and Beyond}

\author{Savas Dimopoulos$^a$ 
and 
Roberto Emparan$^b$\thanks{Also at Depto.\ de F{\'\i}sica
Te\'orica,
Universidad del Pa{\'\i}s Vasco, Bilbao, Spain}}

\address{$^a$ Physics Department, Stanford
          University, Stanford, CA 94305-4060, USA\\
$^b$ 
Theory Division, CERN, CH-1211 Geneva 23, Switzerland}

\setlength{\footnotesep}{0.5\footnotesep}
\maketitle
\begin{abstract}
In string theory, black holes have a minimum mass below which they
transition into highly excited long and jagged strings  --- ``string
balls''. These are the stringy progenitors of black holes; because they
are lighter, in theories of TeV-gravity, they may be more accessible to
the LHC or the VLHC. They share some of the characteristics of black holes,
such as large production cross sections. Furthermore, they evaporate
thermally at the Hagedorn temperature and give rise to
high-multiplicity events containing hard primary photons and charged
leptons, which have negligible standard-model background.
\end{abstract}
%
%\pacs{PACS numbers: }

%
%
% aim for around 700 lines

\newpage

%\section*{}

{\bf Introduction:} An exciting consequence of TeV-scale quantum gravity
\cite{add} is the possibility of producing black holes
(BHs)~\cite{adm,bf,ehm,dl,gt} at the LHC and beyond. Simple estimates of their
production cross section, treating the BHs as general relativistic (GR) objects,
suggest enormous event rates at the LHC  --as large as a BH per sec
\cite{dl,gt}. Furthermore, decays of the BHs into hard primary photons or
charged leptons are a clean signature for detection,  with negligible standard
model background. The production and decay of the experimentally most accessible
light black holes --those with mass $M_{BH}$ near the fundamental Planck scale
$M_P \sim$ TeV-- is clouded by string-theoretic uncertainties. The
purpose of this paper is to discuss these and point out the possible presence 
and properties of light stringy progenitors of the black holes.

According to string theory, the minimum mass $M_{min}$ above which a
black hole can be treated general-relativistically is $M_{min} \sim
{M_s \over g_s^2}$, where $M_s \sim$ TeV is the string scale and $g_s$
is the string coupling --which should be smaller than 1, to trust
string perturbation theory. Therefore, even for moderately small $g_s$,
there is a significant mass range between $M_s$ and $M_{min}$ inside
which the spectrum is intrinsically stringy and the GR approximation
fails. Furthermore, the fundamental Planck scale $M_P$ is typically
smaller than $M_{min}$; so the GR approximation can fail even for
masses larger than $M_P$. For example, if there are $n$ new large and
$m$ small dimensions ($m+n=6$) compactified on an $m$-torus of radius
equal to the string length $l_s$, then $M_P=(2\pi)^{6-n\over 2+n}
g_s^{-2/(2+n)}M_s$; so, $M_P$ is less than $M_{min}$ for any
$n$, if $g_s$ is less than $(2\pi)^{-2/3}\simeq .3$.

An example of a typical value of $g_s$ in theories with large
dimensions is $g_s=2\alpha v_2/(2\pi l_s)^2$ \cite{sht}, where $v_2$ is
the volume contained by the two small dimensions; if $v_2$ is a
two-torus with radius equal to $l_s$, then $g_s$ is between 0.2 and
0.02 depending on which $\alpha$ we choose. Of course, increasing the
size of $v_2$, we can geometrically dilute the value of $\alpha$ and
accommodate larger values of $g_s$. We have nothing to add to the rough
estimates of \cite{dl,gt} in the case where $g_s$ is near 1. For the
rest of the paper we will consider the cases when $g_s^2$ is
significantly less than 1 and there is a large stringy regime between
the scales $M_s$ and $M_{min}$.

{\bf The Correspondence Point:} String theory has provided us with a 
 convincing picture of the evolution
of a black hole at the last stages of its evaporation \cite{BSW,Suss,Ven,hp,dv}. 
As the black hole
shrinks, it eventually reaches the ``correspondence point'' $M_{min}
\sim M_s/g_s^2$, and makes a transition to a configuration dominated by
a highly excited long string. At this point this ``string ball'' is
compact, of size $\sim l_s$. It continues to lose mass by evaporation
at the Hagedorn temperature, and (in six or more space-time dimensions)
at a mass slightly below $M_{min}$ it ``puffs-up'' to a larger
``random-walk'' size $R_{rw}\sim M^{1/2}l_s^{3/2}$. Evaporation, still
at the Hagedorn temperature, then gradually brings the size of the
string ball down towards $l_s$, after which it decays into massless
quanta.

The rationale behind this picture is the following. On  general
grounds, the GR description of a neutral black hole in string theory is
expected to receive large string corrections around the point where the
Schwarzschild radius $R_{BH}$ reaches the string length $l_s$. On the
other hand, both a black hole and a highly excited long string are objects
with a large degeneracy of states. Hence, it is natural to expect that
when the black hole has evaporated down to a size $R_{BH}\sim l_s$
($M\sim M_{min}$) it makes a transition to a configuration dominated by
a highly excited long string. Note again that this may happen before
reaching the Planck scale $M_P$; string corrections can
become important already above the Planck mass.

For this picture to be reliable, the transition at the correspondence
point has to be smooth, at least parametrically. In particular,  the
number of microstates of both objects must be the same at the
transition. The entropy of a long string is proportional to its length,
{\it i.e.} to its mass, $S_{st}\sim M/M_s$, while the Bekenstein
entropy of a black hole is proportional to its area, $S_{BH}\sim
(M/M_{P})^{(n+2)/(n+1)}\sim g_s^{-2}(g_s^2 M/M_S)^{(n+2)/(n+1)}$ (in
$4+n$ dimensions); they match at $M\sim M_{min}$
\cite{BSW,Suss,Ven,hp}. Similarly, the Hagedorn temperature of the
excited string matches the Hawking temperature of a black hole of mass
$M_{min}$. A more subtle point is the comparison between their sizes.
The black hole size at the correspondence point is $\sim l_s$. By
contrast, an excited string has a tendency to spread as a random-walk.
The step size of the random-walk is $l_s$, and its total length
$(M/M_s)l_s$, so the average size of the random-walk, {\it i.e.} the
radius of the string ball, is $R_{rw}\sim (M/M_s)^{1/2}l_s \gg l_s$.
This, however,  neglects the gravitational self-interaction of the
string, which is responsible for keeping the string compact at $\sim
l_s$ near the correspondence point \cite{hp,dv}. For masses below this
point self-gravity becomes less important, and for $n\geq 2$, the rapid
fall-off of gravity very soon makes it insufficient to keep the size of
the string down at $l_s$. So, shortly after the black hole/string ball
transition, the string abruptly puffs-up from string-scale size to
random-walk size \cite{dv}. As we will see, this may have 
observational consequences.

This correspondence picture also suggests that the production cross
section for string balls will match the enormous  black hole cross
section at center of mass energies around $M_s/g_s^2$, but the
transition may involve the effects of strong self-gravity around that
energy.

{\bf String Ball Production:} Let us first calculate the production of
a long, highly excited string from the collision of two light string
states at high $s$ using string perturbation theory. Highly excited
strings are the most likely outcome: due to the exponential degeneracy
of string states at high energies, their phase space volume is much
larger than that of light states with large kinetic energies. The
amplitude for two string states to evolve into a single one, at level
$N\sim s$, is most easily obtained by unitarity from the amplitude for
forward scattering, $A(s,t\to 0)$: the residue at the resonant pole at
$s\sim N$ yields the desired amplitude (squared). Take then the
Virasoro-Shapiro four-point amplitude
\begin{equation}
\label{VSamp}
A(s,t)=2\pi g_s^2{\Gamma(-1-\alpha's/4)\Gamma(-1-\alpha't/4)
\Gamma(-1-\alpha'u/4)
\over \Gamma(2+\alpha's/4)\Gamma(2+\alpha't/4)\Gamma(2+\alpha'u/4)}
\end{equation}
(with $s+t+u=-16/\alpha'$, and $\alpha'=l_s^2$ \cite{stringbooks}).
Near the pole at $\alpha's/4\sim N\gg 1$, in the limit $t\to 0$,
\begin{equation}\label{pole}
A(s,t)\sim 2\pi g_s^2{N^2\over N-\alpha's/4-i\epsilon}+({\mathrm 
terms~analytic~at}~N=\alpha's/4)\,.
\end{equation}
%This pole dominates the amplitude at large $s$, so we
%can equivalently take the limit $s\to\infty, t\to 0$,
%for which
%\begin{equation}
%\label{VSReg}
%A(s,t)\to -g_s^2\alpha'{\pi \over 2t}e^{-i\pi \alpha't/4}
%s^2\,.
%\end{equation}
%Both methods give the same imaginary part. 
The production cross section is
\begin{equation}
\label{ballcs}
%\sigma_{st}={{\mathrm Im}\, A(\alpha's/4=N,t=0)\over s}=
\sigma_{st}={\pi{\mathrm Res}\, A(\alpha's/4=N,t=0)\over s}=
 g_s^2{\pi^2\over 8} \alpha'^2 s\,.
\end{equation}

Note that: (a) although we have used an amplitude with tachyons as
external states, the result should be the same (up to polarization
factors of the initial states) for any colliding particles with rest
mass $\ll \sqrt{s}$. (b) In the limit $s\gg 1/\alpha'$, $t\to 0$, the
amplitude (\ref{VSamp}) has long strings as resonances in the
$s$-channel. Alternatively, the $t$-channel is dominated by the
exchange of a massless closed string over a large distance: the cross
section (\ref{ballcs}) is obtained from single graviton exchange. Since
the latter is universal, it follows that any theory of strings that
contains a graviton will lead, up to numerical factors, to the same
result (\ref{ballcs}). (c) The external states are closed strings. Open
strings can also exchange closed strings, hence gravitons, so the
result above will still be true for them\footnote{$s$-channel
factorization of the amplitude with external open strings leads to the
production of {\it two} long open strings. This doubling will be
unimportant to what follows.}. For open strings there is also
the possibility that the intermediate $t$-channel state is a single
open string, instead of a closed string. The Veneziano amplitude,
appropriate for that case, gives the cross section for production of a
long open string as $\sigma_{ost}=\pi g_o^2 \alpha'$, {\it i.e.} it is
constant. Here $g_o$ is the coupling of open strings. Normally,
$g_o^2\sim g_s$, which implies that this cross section will be
subleading with respect to (\ref{ballcs}) at energies above
$M_s/\sqrt{g_s}$ and, for simplicity, we will ignore it. 
(d) The cross section (\ref{ballcs}) saturates the
unitarity bounds at around $g_s^2 s\alpha'\sim 1$ \cite{acv}. This
implies that the production cross section for string balls grows with
$s$ as in (\ref{ballcs}) only for $M_s< \sqrt{s}\leq M_s/g_s$, while
for $\sqrt{s}>M_s/g_s$ it will remain constant at a value
$\sigma_{st}\sim l_s^2$.\footnote{We thank G.~Veneziano for this
point.} This matches the black hole cross section $\sigma_{BH}\sim
l_s^2(g_s^2 M/M_s)^{2\over n+1}$ at the correspondence point. Conversely,
in light of  the correspondence principle, the computation of the string ball 
cross section at the correspondence point can be viewed as a calculation 
of the black hole production cross section. The agreement with the geometric 
cross section used
in references \cite{bf,dl,gt} may help dispel concerns that the BH production
rate is suppressed \cite{v}.

In summary, the elementary (parton) cross section for string ball/BH
production is
\begin{eqnarray}
\sigma\sim\left\{
\begin{array}{ll}
\displaystyle 
{g_s^2 M_{SB}^2\over M_s^4} &\qquad M_s\ll M_{SB}\leq M_s/ g_s\,,\\
\displaystyle 
{1\over M_s^{2}}&\qquad M_s/ g_s< M_{SB}\leq M_s/ g_s^2\,,\\
\displaystyle
{1\over M_P^{2}}\left({M_{BH}\over M_P}\right)^{2\over n+1} 
&\qquad M_s/ g_s^2<M_{BH}\,.
\end{array}
\right.
\end{eqnarray}
$M_{SB}$ ($M_{BH}$) is the mass of the string ball (black hole), and we
have used $\alpha'=M_s^{-2}$.

The first two mass ranges lead to string balls, the third to black
holes. The cross section interpolates between them and, in all
cases, is large. For example, if $M_s\sim$ TeV, the cross section in
the middle range is $\sim 400$ pb. If there were no kinematical
suppressions, given the LHC luminosity of $30$ fb$^{-1}/$year, this
would give about $10^7$ events per year. However, a string ball is made
out of long jagged string and, consequently, is heavier than $M_s$; to
produce it we need to rely on rare collisions between partons carrying
a significant fraction of the total LHC energy $\sim 14$ TeV. The
calculation of string ball production is quantitatively close to that
of heavy BH production \cite{dl,gt}. From fig.~2 of ref.\ \cite{dl} we
see that  even for a fundamental scale of up to $\sim 4$ TeV, there is a
significant number of BHs with mass $\sim 10$ TeV  produced. We expect the same 
for string balls. As
the ratio of $M_{SB}/M_s$ drops to a few, we may legitimately question
the ``long and jagged string'' picture of a string ball. Here we have
been assuming that the cross section we use to estimate the production
of long strings is adequate for  $M_{SB}/M_s\sim 3$.
Of course, at the VLHC the parameter space to be explored and the
reliability of our estimates will expand significantly.

We do not expect the string balls to be the very first indication of
TeV-gravity; low-lying string states and missing energy through
graviton evaporation are more likely ``discovery'' candidates. String balls 
are interesting because
they are a new form of matter which bridges strings and gravity.

{\bf String Ball Evaporation:} Highly excited long strings (averaged
over degenerate states of the same mass) emit massless particles with a
thermal spectrum at the Hagedorn temperature \cite{ar}. Hence, the
conventional description of evaporation in terms of black body emission
can be applied\footnote{The string ball may also break into two large
pieces by opening two new endpoints on the three-brane.  However, the
results of \cite{lt,abkr} suggest that it will not disintegrate
completely through this mechanism. Then it will evaporate mostly via
thermal radiation.}.  In our case, the emission can take place either
in the bulk (into closed strings) or on the brane (into open strings).
The temperature is the same in both cases,  $T=T_H= {M_s \over
2\sqrt{2}\pi}$ (for type II strings).

For a compact string ball of size $l_s$ there is only this scale in the
problem. Arguing as in \cite{ehm1,ehm} one concludes that the radiation
is distributed roughly equally into all brane and bulk modes. Another
way to understand this \cite{lenny} is to note that the wavelength
$\lambda = {2 \pi \over T_H}$ corresponding to the Hawking temperature
is larger than the size of the black hole. So, the BH (or the
compact string ball) is, to first approximation, a point-radiator and, consequently,
 emits mostly $s$-waves. This indicates that it decays equally
to a particle on the brane and in the bulk, since s-wave emission is sensitive
only to the radial coordinate and does not make use of the extra angular
modes available in the bulk. Because there are many more species of particles ($\sim
60$) 
on our
brane than in the bulk, this has the crucial consequence that the
radiator decays visibly to standard model (SM)
particles~\cite{ehm,lenny}.

However, when the string ball puffs-up to the larger random-walk size,
its spatial extent can approach or exceed the wavelength of the emitted
quanta, which implies that it can use more of the higher angular modes
that the additional dimensions provide. We can estimate how many of
these will become available. First note that the radius of the string
ball has increased to $l_s\sqrt{M/M_s}$, which near the correspondence
point (when the string ball is largest) is $l_s/g_s$. A black body
 of radius $R$ emits 
modes with angular momentum $l\leq l_{max}=\omega R$,  where
$\omega=T_H=M_s/2\sqrt{2}\pi$ is the frequency of the radiation. So for
$R=l_s/g_s$ we get $l_{max}={1/2\sqrt{2} \pi g_s}$; for
$g_s$ around $0.1$  this means $l=1$ or $2$. In $4+n$ dimensions the
degeneracy of angular momentum modes is ${2l+n+1\over n+1}{l+n\choose
l}$ compared to the degeneracy $2l+1$ in 4 dimensions. For $n=2$ and
$l=1$ $(2)$ the relative enhancement for the bulk modes is $5/3$
$(33/8)$, which implies an increase of the radiation into the bulk, but
still outnumbered by the brane modes. The same conclusion follows from
the black body radiation formula $dE_D/dt\sim A_DT^D$ ($A_D$ is the
area of the radiator in $D$ dimensions) \cite{ehm,ehm1}: for a
random-walk-size string ball at the Hagedorn temperature the relative
enhancement of bulk radiation is $(dE_{4+n}/dt)/ (dE_4/dt) \sim
(2\sqrt{2}\pi g_s)^{-n}$, which is again of order one for $n=2$ and
$g_s\sim 0.1$. With more extra dimensions, larger angular momentum,
 and smaller $g_s$ the above formulas give a rapid relative 
 increase in the number of bulk
modes, which eventually dominate the radiation. This, however, is a
temporary effect: as the string ball decays, its size shrinks towards
$l_s$ and, once again, it becomes a small radiator emitting mostly on the brane.

A string ball of mass $M$ will decay, on the average, into roughly  $M
/ T_H$ particles of mean energy $\sim T_H$.  Just like BHs,   the
ensemble of string balls  should decay about equally  to
each of the $\approx 60$ particles of the SM. Since there are  six
charged leptons  and one photon, we expect $\sim 10\%$  of the
particles to be hard, primary charged leptons and $\sim 2\%$ of the
particles to be hard photons, each  carrying hundreds of GeV of energy.
This is a  clean signal, with  negligible background, as the production
of  leptons or photons through SM processes in  high-multiplicity
events at the LHC occurs at a much smaller rate than  the string ball
production. These events are also easy to  trigger on, since they
contain at least one prompt lepton or photon with energy above 100 GeV,
as well as energetic quark and gluon jets. Measuring the mean energy of
the decay products determines the Hagedorn temperature and,
consequently, the string mass scale $M_s$. 

The fraction of missing energy
in string ball events could be an interesting probe of TeV-gravity
physics. The  neutrinos' contribution is small, since they account for
just  $\sim 5\%$ of the final particles. On the other hand, significant
amounts of missing energy  --resulting from bulk emission from a
puffed-up string  ball-- could signal the presence of several large new
dimensions.

\begin{table}[tb]
\begin{center}
\caption{New particles and their associated mass scales. Typically,
$M_s<M_P<M_s/g_s^2$.}
\medskip
\begin{tabular}{lc}
%\hline
~Particles       		& Mass Scale   		\\
\hline
1.~Higher-dimensional graviton	& $M_P$  			\\
2.~Low-lying string excitations	& $M_s$  			\\
3.~String Balls    		& $M_s\ll E\leq M_s/g_s^2$  	\\
4.~Black Holes   		& $E>M_s/g_s^2$ 		\\
%\hline
\end{tabular}
\end{center}
\label{table}
%\vspace*{-0.3in}
\end{table}

{\bf Conclusions:} TeV-gravity theories have at least four new types of
particles and three associated mass scales, as shown in the table. If
$g_s \sim 1$, the mass scales coincide and calculability is lost; BHs
are expected to dominate the dynamics above $M_s$. If $g_s^2 \ll 1$,
then there is a separation between the mass scales and we expect to
probe the physics of the particles roughly in the order 1, 2, 3, 4. In
this case LHC may be able  to probe the physics of string balls, but is
less likely to produce BHs. Higher energy colliders, such as the VLHC,
will have a better chance of studying BHs. Through their evaporation,
black holes will evolve into string balls and, eventually, into
low-lying string states, giving us a glimpse of all the stages of this
exciting physics.

%\vskip 2cm

\bigbreak\bigskip\bigskip\centerline{{\bf
Acknowledgements}}\nobreak 

\vskip .5cm

We would like to thank Enrique \'Alvarez, Jos\'e L.~F.\ Barb\'on and 
Gabriele Veneziano for valuable conversations. The work of SD is
supported by the NSF grant PHY-9870115 at Stanford University. The work
of RE is partially supported by UPV grant 063.310-EB187/98 and CICYT
AEN99-0315. SD thanks the theory group at CERN for its hospitality.

%\newpage

\end{document}